\documentclass[letterpaper,twocolumn,english,superscriptaddress,prl,aps,showpacs]{revtex4}
\usepackage[T1]{fontenc}
\usepackage[latin1]{inputenc}
\usepackage{amsmath}
\usepackage{graphicx}
\usepackage{amssymb}

\makeatletter
\@ifundefined{textcolor}{}
{%
 \definecolor{BLACK}{gray}{0}
 \definecolor{WHITE}{gray}{1}
 \definecolor{RED}{rgb}{1,0,0}
 \definecolor{GREEN}{rgb}{0,1,0}
 \definecolor{BLUE}{rgb}{0,0,1}
 \definecolor{CYAN}{cmyk}{1,0,0,0}
 \definecolor{MAGENTA}{cmyk}{0,1,0,0}
 \definecolor{YELLOW}{cmyk}{0,0,1,0}
 }

\@ifundefined{definecolor}{\@ifundefined{definecolor}
 {\@ifundefined{definecolor}
 {\usepackage{color}}{}
}{}
}{}\makeatother

\makeatother

\usepackage{babel}
\DeclareMathAlphabet{\mathpzc}{OT1}{pzc}{m}{it}

\makeatother

\usepackage{babel}

\begin{document}

\author{Alexey A. Kovalev}

\affiliation{Department of Physics and Astronomy, University of California, Los
Angeles, California 90095, USA}

\author{Jairo Sinova}

\affiliation{Department of Physics, Texas A\&M University, College Station, TX
77843-4242, USA}

\affiliation{Institute of Physics ASCR, Cukrovarnická 10, 162 53 Praha 6, Czech
Republic }

\author{Yaroslav Tserkovnyak}

\affiliation{Department of Physics and Astronomy, University of California, Los
Angeles, California 90095, USA}

\title{Anomalous Hall Effect in Disordered Multi-band Metals}
\begin{abstract}
We present a microscopic theory of the anomalous Hall effect in metallic
multi-band ferromagnets, which accounts for all %disorder-strength
scattering-independent contributions, i.e., both the intrinsic and
the so-called side jump. For a model of Gaussian disorder, the anomalous
Hall effect is expressed solely in terms of the electronic band structure
of the host material. Our theory handles systematically the interband-scattering
coherence effects. We demonstrate the method in the two-dimensional
Rashba and three-dimensional ferromagnetic (III,Mn)V semiconductor
models. Our formalism is directly amenable to \textit{ab initio} treatments
for a wide range of ferromagnetic metals. 
\end{abstract}

\date{\today{}}

\pacs{72.15.Eb, 72.20.Dp, 72.20.My, 72.25.-b}

\maketitle
\textit{Introduction.}---While the anomalous Hall effect (AHE) has
attracted generous attention from the physics community starting with
the seminal work by Karplus and Luttinger \citep{Karplus:sep1954},
its full theoretical understanding remains incomplete \citep{Nagaosa:May2010}.
Contributions to the AHE in ferromagnetic metals can be separated
according to the dependence on the quasiparticle transport life time
$\tau$, e.g., $\sigma_{AH}(\tau)=\sigma^{(0)}_{AH}+\alpha_{H}\sigma_{xx}(\tau)+\dots$,
where $\sigma^{(0)}_{AH}$ is the scattering-independent contribution,
and $\alpha_{H}\sigma_{xx}(\tau)$, %is the scattering-independent coefficient called Hall angle corresponding to the skew scattering 
%and $\sigma_{xx}\sim\tau$ is the diagonal conductivity. 
usually termed skew scattering, is linear in $\tau$ in the Drude
limit (i.e., $\tau\omega_{F}\gg1$, where $\hbar\omega_{F}$ is the Fermi energy).
The scattering-independent term $\sigma^{(0)}_{AH}$ is usually further
separated into the intrinsic contribution (IC), $\sigma^{{\rm int}}_{AH}$,
and the side-jump contribution (SJC) $\sigma^{{\rm sj}}_{AH}\equiv\sigma^{(0)}_{AH}-\sigma^{{\rm int}}_{AH}$.
$\sigma^{{\rm int}}_{AH}$ is defined as the extrapolation of the
ac Hall conductivity to zero frequency in a clean system, with the
limit $\tau^{-1}\rightarrow0$ taken before $\omega\rightarrow0$
\citep{Nagaosa:May2010}. The IC has been shown to
be linked to the Berry phase of the spin-orbit (SO) coupled Bloch
electrons \citep{Sundaram:jun1999}. It is the most directly calculable
contribution to the AHE in ferromagnetic semiconductors, transition metals and complex oxides \citep{Jungwirth:may2002,Mathieu:Jun2004}.

A wide range of strongly SO coupled ferromagnetic metals exhibit scattering-independent
$\sigma^{(0)}_{AH}$ in the $\sigma_{AH}$ signal with a sizable deviation
from the calculated $\sigma^{{\rm int}}_{AH}$ \citep{Jungwirth:may2002,Nagaosa:May2010,Mathieu:Jun2004},
which implies substantial SJC. Although an experimental separation
of IC and SJC has been suggested by studying an interplay of different
kinds of disorder (e.g., phonons and impurities) at finite temperatures
\citep{Tian:Aug2009}, comparison with the theoretical expectation
for $\sigma^{(0)}_{AH}$ has been hampered by the lack of a simple rigorous
formalism that would allow a reliable calculation of $\sigma^{{\rm sj}}_{AH}$
in complex multi-band systems. It is thus desirable to develop a general
procedure for calculating all scattering-independent contributions
to allow for a systematic comparison with experiments and engineering
of materials with necessary AHE properties. It should be possible
to identify the SJC by ac measurements in the clean limit (i.e., $\tau^{-1}<\Delta$,
the characteristic SO band-energy splitting): The AHE is modified by the effects of
disorder at low frequencies, while at intermediate frequencies, $\tau^{-1}<\omega<\Delta$,
the IC should be recovered as interband coherences caused by disorder scattering
do not build up \citep{Inoue:jul2004}.

In this Letter, we calculate the AHE in metallic noninteracting multi-band
systems in the presence of delta-correlated Gaussian disorder, expressing
the final result solely through the Bloch wave functions, similar
to the theory of the intrinsic AHE \citep{Sundaram:jun1999}. There
is no skew-scattering contribution ($\propto\sigma_{xx}$) within
such model of disorder, and we assume nondegenerate bands. The main results of this Letter for the IC
and SJC are given in Eqs.~(\ref{iAHE})-(\ref{Skew scattering})
requiring only the material's electronic band structure as the input.
These equations should apply to a wide range of metallic materials
exhibiting scattering-independent AHE \citep{Nagaosa:May2010}.
The present theory has been tested on the two-dimensional (2D) Rashba
Hamiltonian reproducing known results \citep{Sinitsyn:jan2007,Kovalev:MAY2009}.
Furthermore, the SJC is found to dominate the AHE in a model of metallic
ferromagnetic (III,Mn)V semiconductors.

The Berry phase of Bloch states has a significant effect on transport
properties, particularly on the AHE. The origin of this lies in the
anomalous velocity proportional to the external electric field that
modifies the group velocity \citep{Sundaram:jun1999}, i.e., $\hbar\dot{\mathbf{r}}=\partial_{\mathbf{k}}\varepsilon_{\eta}(\mathbf{k})-e\mathbf{E}\times\boldsymbol{\mathcal{B}}_{\eta}(\mathbf{k})$,
where $\varepsilon_{\eta}(\mathbf{k})$ is the band energy, $\mathbf{E}$
external electric field, $\boldsymbol{\mathcal{B}}_{\eta}(\mathbf{k})=i\partial_{\mathbf{k}}\times\langle u_{\eta}|\partial_{\mathbf{k}}|u_{\eta}\rangle$
Berry curvature, and $e$ particle charge ($e<0$ for electrons).
Modifications to the motion of electrons (holes) in the $\eta$th
band are defined solely in terms of the periodic part
of the Bloch wave functions $|u_{\eta}(\mathbf{k})\rangle$. Below,
we will show that this is no longer the case due to band mixing,
in the presence of an even infinitesimally small disorder.

\textit{IC and SJC from the band structure.}---Consider a general
multi-band noninteracting system, in the $\mathbf{k}\cdot\mathbf{p}$ expansion, up to some order in $\mathbf{k}$
about an extremum point in the Brillouin zone. Below, we
will consider Luttinger and Rashba Hamiltonians as particular realizations
of such expansions. In the position representation, our $N$-band projected
Hamiltonian is expressed via ``envelope fields'':\begin{equation}
\mathcal{H}_{0}=\sum_{\eta,\eta'}\int dr\Psi_{\eta}^{\dagger}(r)[\hat{H}_{0}(-i\boldsymbol{\nabla}_{r})]_{\eta\eta'}\Psi_{\eta'}^{\dagger}\;.\label{Hamiltonian}\end{equation}
 Here, $\Psi_{\eta}$ is the ``envelope field'' of the $\eta$th
band, with index $\eta$ running from $1$ to $N$.
We suppose that all information about our system, such as SO interaction
or exchange field, is contained in the matrix structure of $\hat{H}_{0}(\mathbf{k})$,
where $\mathbf{k}$ corresponds to $-i\boldsymbol{\nabla}_{r}$. The
phenomenological exchange field is introduced within
the framework of a mean-field description. In addition to the band
Hamiltonian, we include a scalar delta-correlated Gaussian disorder $V(\mathbf{r})$
with $\left\langle V(\mathbf{r})V(\mathbf{r}')\right\rangle =\hbar^{2}\mathcal{V}\delta(\mathbf{r}-\mathbf{r}')$.

In the absence of disorder, the anomalous velocity mentioned above leads to the intrinsic spin Hall conductance
\begin{equation}
\sigma_{ij}^{\rm int}=\dfrac{e^{2}}{\hbar}\sum_{\eta}\int\frac{d^{n}k}{(2\pi)^{n}}\frac{d\omega}{2\pi}n_{F}i[\mathcal{\hat{A}}_{k_{i}}\mathcal{\hat{A}}_{k_{j}}-\mathcal{\hat{A}}_{k_{j}}\mathcal{\hat{A}}_{k_{i}}]_{\eta\eta}A_{\eta}\;,\label{iAHE}\end{equation}
 where $n=2$ or $3$ is the number of spatial dimensions, $n_{F}(\omega)$
is the Fermi distribution function, and $A_{\eta}(\mathbf{k},\omega)$ is the spectral
function of the $\eta$th band. The anomalous transport is governed by the Berry-connection matrix $\mathcal{\hat{A}}_{\mathbf{k}}=i\hat{U}^{\dagger}\partial_{\mathbf{k}}\hat{U}$, where $\hat{U}$ is a $\mathbf{k}$-dependent unitary matrix transforming the $\mathbf{k}\cdot\mathbf{p}$ Hamiltonian into a diagonal band-energy matrix $\hat{\varepsilon}(\mathbf{k})=\hat{U}^{\dagger}\hat{H}_{0}(\mathbf{k})\hat{U}$.

\begin{figure}
\includegraphics[clip,width=1\columnwidth]{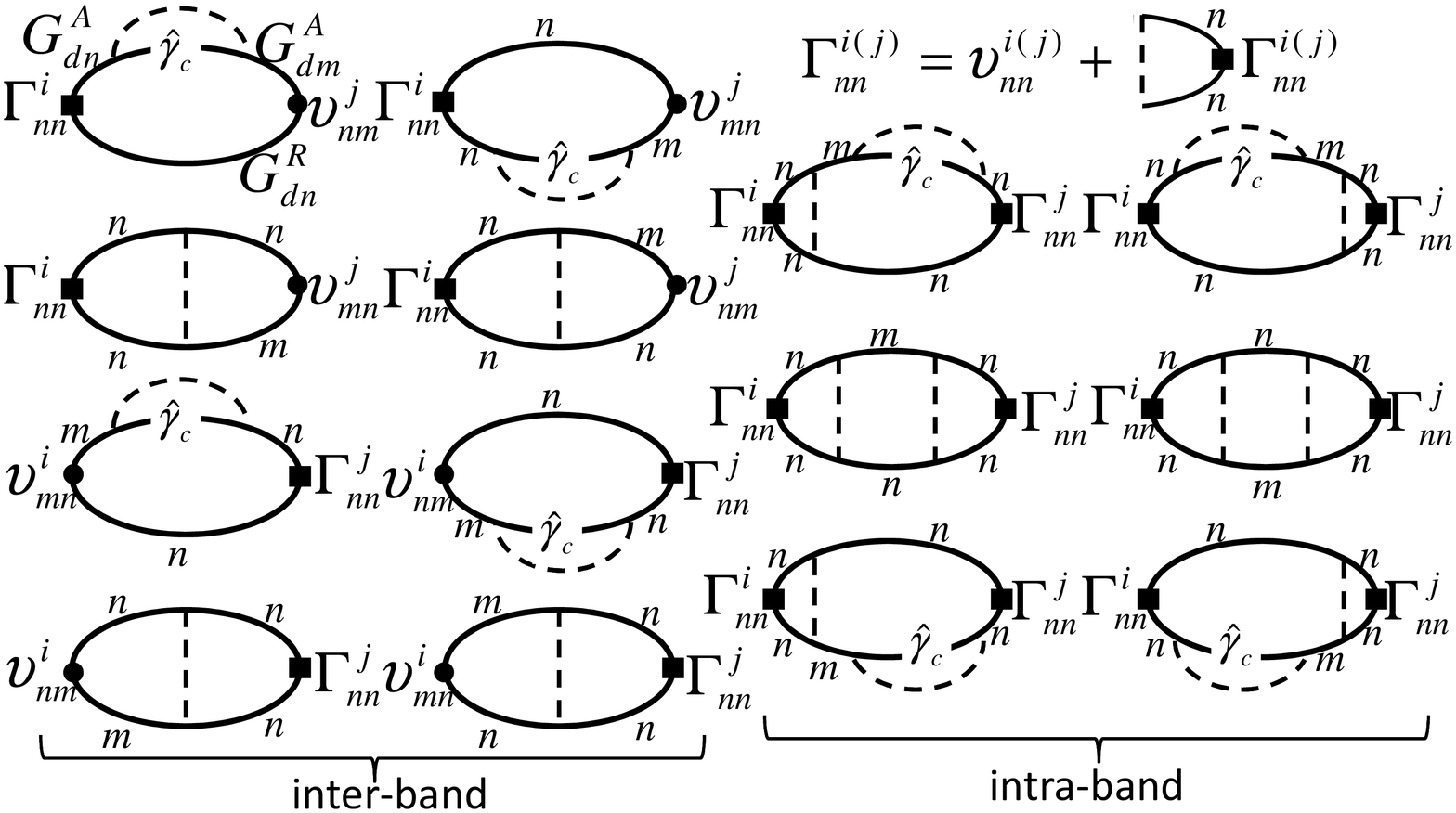} 
\caption{The SJC diagrams for $\sigma_{ij}$ expressed in the band eigenstate basis described
by the indices $m$ and $n$ ($m\neq n$). The bold lines correspond
to $\hat{G}_{d}^{R(A)}$ (that can be replaced by disorder-free GF's
for calculation of the SJC \citep{Sinitsyn:jan2007}) while dashed lines correspond to disorder
strength $\mathcal{V}$. By iterating Eq.~(\ref{DensityEq1}) and expanding
as in Eq.~(\ref{Expansion}), the leading-order contributions in Eq.~(\ref{CurentI-Calc}) can be expressed as two components of SJC. The
diagrams beginning and ending with velocity involving single (multiple)
band(s) are termed as intra(inter)-band diagrams.}
\label{fig1} 
\end{figure}

In this Letter, we show that the SJC contains two terms
expressed via the electronic band structure as
\begin{align}
\sigma_{ij}^{a}=&\dfrac{e^{2}}{\hbar}\sum_{\eta=1}^{N}\int\frac{d^{n-1}k_{\eta}}{(2\pi)^{n}}\dfrac{1}{\left|\partial_{\hbar\mathbf{k}}\varepsilon_{\eta}\right|[\hat{\gamma}_{c}]_{\eta\eta}}\nonumber\\
&\times{\rm Tr}\left\{\left([\hat{\gamma}_{c}]_{\eta\eta}\hat{U}\mathcal{\hat{A}}_{k_{i}}\hat{S}_{\eta}\hat{U}^{\dagger}-\hat{U}\hat{S}_{\eta}\mathcal{\hat{A}}_{k_{i}}\hat{\gamma}_{c}\hat{S}_{\eta}\hat{U}^{\dagger}\right)\hat{P}_{j}\right.\nonumber\\
&+\left.(\partial_{\hbar k_{j}}\varepsilon_{\eta})\hat{S}_{\eta}\mathcal{\hat{A}}_{k_{i}}(\hat{1}-\hat{S}_{\eta})\hat{\gamma}_{c}\right\}+{\rm c.c.\,,}
\label{Side-jump}\\
\sigma_{ij}^{b}=&e^{2}{\displaystyle \sum_{\eta=1}^{N}}\int\frac{d^{n-1}k_{\eta}}{(2\pi)^{n}}\dfrac{i}{2\left|\partial_{\hbar\mathbf{k}}\varepsilon_{\eta}\right|[\hat{\gamma}_{c}]_{\eta\eta}}\nonumber\\
&\times{\rm Tr}\left\{\hat{U}\hat{S}_{\eta}\hat{U}^{\dagger}\hat{P}_{j}\hat{U}\hat{S}_{\eta}\hat{\gamma}_{c}\hat{C}_{\eta}\hat{U}^{\dagger}\hat{P}_{i}\right.\nonumber\\
&-\hat{U}\hat{S}_{\eta}\hat{U}^{\dagger}\hat{P}_{i}\hat{U}\hat{S}_{\eta}\hat{\gamma}_{c}\hat{C}_{\eta}\hat{U}^{\dagger}\hat{P}_{j}\nonumber\\
&+[\hat{\gamma}_{c}]_{\eta\eta}\hat{U}\hat{S}_{\eta}\hat{U}^{\dagger}\hat{P}_{i}\hat{U}\hat{C}_{\eta}\hat{U}^{\dagger}\hat{P}_{j}\nonumber\\
&+\left.2(\partial_{\hbar k_{i}}\varepsilon_{\eta})\hat{U}\hat{S}_{\eta}\hat{\gamma}_{c}\hat{C}_{\eta}\hat{U}^{\dagger}\hat{P}_{j}\right\}+{\rm c.c.}\;.\label{Skew scattering}\end{align} Here, $[\hat{S}_{\eta}]_{ij}=\delta_{ij}\delta_{i\eta}$ ($\delta_{ij}$ is the Kronecker delta symbol) and $[\hat{C}_{\eta}]_{ij}=(\epsilon_{\eta}-\epsilon_{i})^{-1}\delta_{ij}$ for $i\neq\eta$ and zero otherwise.
\begin{equation}
\hat{\gamma}_{c}(\mathbf{k}')=\hat{U}(\mathbf{k}')^{\dagger}\left({\displaystyle \sum_{\eta=1}^{N}}\int{\displaystyle \frac{d^{n-1}k_{\eta}}{(2\pi)^{n-1}}\dfrac{\hat{U}(\mathbf{k})\hat{S}_{\eta}\hat{U}(\mathbf{k})^{\dagger}}{2\left|\partial_{\hbar\mathbf{k}}\varepsilon_{\eta}\right|}}\right)\hat{U}(\mathbf{k}')\;,\label{SE1}\end{equation}
 and the matrix $\mathbf{\hat{P}}$ corresponding to a subset of vertex corrections denoted by $\hat{\boldsymbol{\Gamma}}$ in Fig. \ref{fig1}
is defined by a total of $N^{2}$ linear equations with $N$ equations
\begin{equation}
\mathbf{\hat{P}}=\int{\displaystyle \frac{d^{n-1}k_{\eta}}{(2\pi)^{n-1}}\dfrac{\hat{U}\hat{S}_{\eta}\hat{U}^{\dagger}\mathbf{\hat{P}}\hat{U}\hat{S}_{\eta}\hat{U}^{\dagger}-(\partial_{\hbar\mathbf{k}}\varepsilon_{\eta})\hat{U}\hat{S}_{\eta}\hat{U}^{\dagger}}{2\left|\partial_{\hbar\mathbf{k}}\varepsilon_{\eta}\right|[\hat{\gamma}_{c}]_{\eta\eta}}}\label{Pmatrix}\end{equation}
for each $\eta$. In the above equations, $d^{n-1}k_{\eta}$ stands for the integration
over the Fermi surface of the $\eta$th band. The SJCs
in Eqs.~(\ref{Side-jump}) and (\ref{Skew scattering}) are distinct
from the diagrammatic point of view as will be clear below. The
mechanism of the former SJC relies on the effects
related to the Berry curvature hence the dependence on $\mathcal{\hat{A}}_{\mathbf{k}}$.

\textit{Derivation.}---In various theories of the AHE in multi-band
systems, it is common to express the conductivity via the Green's functions (GF's) calculated
in equilibrium \cite{Kovalev:MAY2009}. Such description
requires as input information about both the disorder and
band structure. By taking advantage of the band eigenstate representation,
we will express our results only via the band structure. To fulfill
this, we first express all GF's via their diagonal parts as\begin{align}
\hat{G}_{c}^{R(A)}&=\hat{U}^{\dagger}\hat{G}_{{\rm eq}}^{R(A)}\hat{U}=\left[1-\hat{G}_{d}^{R(A)}\hat{\Sigma}_{{\rm nd}}^{R(A)}\right]^{-1}\hat{G}_{d}^{R(A)}\nonumber\\
&={\textstyle \hat{G}_{d}^{R(A)}}+{\textstyle \hat{G}_{d}^{R(A)}}\hat{\Sigma}_{{\rm nd}}^{R(A)}{\textstyle \hat{G}_{d}^{R(A)}}+\dots\;.\label{Expansion}\end{align}
Here and henceforth, the eigenstate representation
is denoted by index $c$, $\hat{\Sigma}_{c}^{R(A)}=\hat{U}^{\dagger}\hat{\Sigma}_{{\rm eq}}^{R(A)}\hat{U}=\hat{\Sigma}_{d}^{R(A)}+\hat{\Sigma}_{{\rm nd}}^{R(A)}$
is the self-energy matrix separated into the diagonal and off-diagonal
parts, ${\textstyle \hat{G}_{{\rm eq}}^{R(A)}}=\hbar(\hbar\omega-\hat{H}_{0}-\hat{\Sigma}_{{\rm eq}}^{R(A)})^{-1}$
is the retarded (advanced) GF in equilibrium and
${\textstyle \hat{G}_{d}^{R(A)}}=\hbar(\hbar\omega-\hat{H}_{0}-\hat{\Sigma}_{d}^{R(A)})^{-1}$
is the corresponding diagonal GF. The imaginary part of ${\textstyle \hat{G}_{d}^{R(A)}}$
proportional to the spectral function $\hat{A}=i(\hat{G}_{d}^{R}-\hat{G}_{d}^{A})$
will be integrated out reducing the problem to Fermi-surface integration.
It is crucial to keep off-diagonal matrices $\hat{\Sigma}_{{\rm nd}}^{R(A)}$
in Eq.~(\ref{Expansion}) up to the necessary order (the first order
for the leading-order AHE) since they contain information about the
interband coherences that play an important role in the AHE.

Our starting point is the expressions for the current densities derived within the Kubo-Streda
formalism \citep{Streda:aug1982} by summing all noncrossed diagrams.
These expressions are also obtained in \citep{Kovalev:MAY2009} using the Keldysh formalism: \begin{align}
j_{i}^{\rm I}= & -\dfrac{e^{2}}{\hbar}\int\frac{d^{n}k}{(2\pi)^{n}}\frac{d\omega}{2\pi}\mathbf{E}\partial_{\omega}n_{F}{\rm Tr}\left[\mathcal{V}\hat{G}_{{\rm eq}}^{R}\boldsymbol{\hat{\rho}}\hat{G}_{{\rm eq}}^{A}\hat{\upsilon}_{i}\right.\nonumber \\
 & \left.+\left(\hat{G}_{{\rm eq}}^{R}\boldsymbol{\hat{\upsilon}}\hat{G}_{{\rm eq}}^{A}-(\hat{G}_{{\rm eq}}^{A}\boldsymbol{\hat{\upsilon}}\hat{G}_{{\rm eq}}^{A}+\hat{G}_{{\rm eq}}^{R}\boldsymbol{\hat{\upsilon}}\hat{G}_{{\rm eq}}^{R})/2\right)\hat{\upsilon}_{i}\right],\label{CurentI-Calc}\\
j_{i}^{\rm II}= & \dfrac{e^{2}}{\hbar}\int\frac{d^{n}k}{(2\pi)^{n}}\frac{d\omega}{2\pi}\mathbf{E}n_{F}{\rm Tr}\left[\mathcal{V}\hat{G}_{{\rm eq}}^{R}\boldsymbol{\hat{\rho}}_{E}^{R}\hat{G}_{{\rm eq}}^{R}\hat{\upsilon}_{i}\right.\nonumber \\
 & +\left(\hat{G}_{{\rm eq}}^{R}\boldsymbol{\hat{\upsilon}}\partial_{\omega}\hat{G}_{{\rm eq}}^{R}-\partial_{\omega}\hat{G}_{{\rm eq}}^{R}\boldsymbol{\hat{\upsilon}}\hat{G}_{{\rm eq}}^{R}\right)\hat{\upsilon}_{i}/2\Bigr]+{\rm c.c.}\;,\label{CurrentII-Calc}\end{align}
where the vector-valued matrices $\boldsymbol{\hat{\rho}}(\omega)$
and $\boldsymbol{\hat{\rho}}_{E}^{R}(\omega)$ satisfy the following equations ($\boldsymbol{\hat{\upsilon}}=\partial_{\hbar\mathbf{k}}\hat{H}_{0}$):
\begin{align}
\boldsymbol{\hat{\rho}}= & \int\frac{d^{n}k}{(2\pi)^{n}}\left(\mathcal{V}\hat{G}_{{\rm eq}}^{R}\boldsymbol{\hat{\rho}}\hat{G}_{{\rm eq}}^{A}+\hat{G}_{{\rm eq}}^{R}\boldsymbol{\hat{\upsilon}}\hat{G}_{{\rm eq}}^{A}\right)\;,\label{DensityEq1}\\
\boldsymbol{\hat{\rho}}_{E}^{R}= & \int\frac{d^{n}k}{(2\pi)^{n}}\left[\mathcal{V}\hat{G}_{{\rm eq}}^{R}\boldsymbol{\hat{\rho}}_{E}^{R}\hat{G}_{{\rm eq}}^{R}\right.\label{DensityEq2}\nonumber\\
 & \left.+\left(\hat{G}_{{\rm eq}}^{R}\boldsymbol{\hat{\upsilon}}\partial_{\omega}\hat{G}_{{\rm eq}}^{R}-\partial_{\omega}\hat{G}_{{\rm eq}}^{R}\boldsymbol{\hat{\upsilon}}\hat{G}_{{\rm eq}}^{R}\right)/2\right]\;.\end{align}

The equilibrium GF's can be found by using the self-energy, $\hat{\Sigma}_{{\rm eq}}^{R(A)}(\omega)=\hbar \mathcal{V}\int{\displaystyle {d^{n}k}{\textstyle \hat{G}_{{\rm eq}}^{R(A)}}(\mathbf{k},\omega)/{(2\pi)^{n}}},$
where only the imaginary part of $\hat{\Sigma}_{{\rm eq}}^{R(A)}$
should be calculated since the real part can be combined with the
Hamiltonian $\hat{H}_{0}$ introducing corrections to the eigenstates
and eigenenergies of $\hat{H}_{0}$ that vanish as we take the strength
of disorder to zero. Using Eq.~(\ref{Expansion}), we can write ${\rm Im}\hat{\Sigma}_{{\rm eq}}^{R(A)}=\mp\hbar \mathcal{V}\hat{\gamma}$
up to the lowest order in $\mathcal{V}$ where \begin{equation}
\hat{\gamma}(\omega)={\displaystyle \sum_{\eta=1}^{N}}\int{\displaystyle \frac{d^{n-1}k_{\eta}}{(2\pi)^{n-1}}\dfrac{\hat{U}\hat{S}_{\eta}\hat{U}^{\dagger}}{2\left|\partial_{\hbar\mathbf{k}}\varepsilon_{\eta}\right|}}\label{Self Energy}\end{equation}
is determined solely by the
electronic band structure, and the integral runs over the wave-vector surface
corresponding to energy $\hbar\omega$ (we will only need
$\hat{\gamma}$ at the Fermi level).

We next rewrite Eqs.~(\ref{CurentI-Calc}) and (\ref{CurrentII-Calc})
through GF's ${\textstyle \hat{G}_{d}^{R(A)}}$ using Eq.~(\ref{Expansion}). In the limit of vanishing disorder, the first
term corresponding to vertex corrections in Eq.~(\ref{CurrentII-Calc})
vanishes. In the remaining terms, it is sufficient to use the diagonal GF's ${\textstyle \hat{G}_{d}^{R(A)}}$
instead of ${\textstyle \hat{G}_{{\rm eq}}^{R(A)}}$. We can identify the IC by combining $j^{\rm II}$ with several terms from $j^{\rm I}$:
\begin{align}
j_{i}^{\rm int}&=\dfrac{e^{2}}{2\hbar}\mathbf{E}{\displaystyle \int}{\displaystyle \frac{d^{n}k}{(2\pi)^{n}}}{\displaystyle \frac{d\omega}{2\pi}}\nonumber\\
&\hspace{-0.7cm}\times\left\{ n_{F}{\rm Tr}\Bigl[\left(\hat{G}_{d}^{R}\boldsymbol{\hat{\upsilon}}_{c}\partial_{\omega}\hat{G}_{d}^{R}-\partial_{\omega}\hat{G}_{d}^{R}\boldsymbol{\hat{\upsilon}}_{c}\hat{G}_{d}^{R}\right)(\hat{\upsilon}_{c})_{i}+{\rm c.c.}\Bigr]\right.\nonumber\\
&\hspace{-0.7cm}-\left.\partial_{\omega}n_{F}{\rm Tr}\left[\left(2\hat{G}_{d}^{R}\boldsymbol{\hat{\upsilon}}_{c}\hat{G}_{d}^{A}-\hat{G}_{d}^{A}\boldsymbol{\hat{\upsilon}}_{c}\hat{G}_{d}^{A}-\hat{G}_{d}^{R}\boldsymbol{\hat{\upsilon}}_{c}\hat{G}_{d}^{R}\right)(\hat{\upsilon}_{c})_{i}\right]\right\}.\nonumber\end{align}
Using integration by parts and keeping only zeroth-order terms in
$\mathcal{V}$, we arrive at Eq.~(\ref{iAHE}).

To obtain the remaining terms in Eq.~(\ref{CurentI-Calc}) up to the zeroth order in $\mathcal{V}$ we expand Eq.~(\ref{DensityEq1}) into an infinite series, furthermore substituting this series into Eq.~(\ref{CurentI-Calc}). In the band eigenstate representation, we can further replace the GF's via diagonal ones according to Eq.~(\ref{Expansion}).
The resulting infinite sum has the terms of order $\mathcal{V}^{-1}$:\[
\sigma_{ij}=\dfrac{e^{2}}{\hbar \mathcal{V}}{\displaystyle \sum_{\eta=1}^{N}}\int{\displaystyle \frac{d^{n-1}k_{\eta}}{(2\pi)^{n}}\partial_{\hbar k_{i}}\varepsilon_{\eta}\dfrac{\partial_{\hbar k_{j}}\varepsilon_{\eta}-[\hat{U}^{\dagger}\hat{P}_{j}\hat{U}]_{\eta\eta}}{2\left|\partial_{\hbar\mathbf{k}}\varepsilon_{\eta}\right|[\hat{\gamma}_{c}]_{\eta\eta}}}\;,\]
leading to the symmetric part of the conductivity, which describes the anisotropic magnetoresistance. The more interesting terms contributing to the AHE appear at zeroth order in $\mathcal{V}$ and can be graphically represented as two sets of diagrams (see Fig. \ref{fig1}). The inter-band diagrams, corresponding to calculating $\boldsymbol{\hat{\rho}}=\mathbf{\hat{P}}/\mathcal{V}+\mathcal{O}(\mathcal{V})$ in Eq.~(\ref{CurentI-Calc}) up to the most singular (i.e., $\mathcal{V}^{-1}$) order, lead to Eq.~(\ref{Side-jump}). The intra-band diagrams, corresponding to calculating $\boldsymbol{\hat{\rho}}$ in Eq.~(\ref{CurentI-Calc}) up to the zeroth order in $\mathcal{V}$, lead to Eq.~(\ref{Skew scattering}). Here, $\mathbf{\hat{P}}$ is an $N\times N$ matrix
given by the solution to Eq.~(\ref{Pmatrix}), which corresponds to the leading-order vertex correction to the bare velocity captured by $\hat{\boldsymbol{\Gamma}}$. 

\textit{Application to Rashba and Luttinger models.}---We first apply
Eqs.~(\ref{iAHE})-(\ref{Skew scattering})
to a Rashba ferromagnet with $\{-\Omega;\Omega\}$ band gap
at $k=0$ arriving at expressions (Table I in supplementary
material \citep{supp}) consistent with the previous works \citep{Sinitsyn:jan2007,Kovalev:MAY2009}.
The vertex corrections lead to important contributions and should
in general be considered.

However, for inversion-symmetric systems with $\hat{H}_{0}(\mathbf{k})=\hat{H}_{0}(-\mathbf{k})$,
\textit{the vertex corrections vanish} for short-ranged disorder as
can be seen by inspecting the $\boldsymbol{\hat{P}}$ independent
term in Eq.~(\ref{Pmatrix}). Similar vanishing of the vertex corrections
takes place in calculations of the anisotropic magnetoresistance and
SHE \citep{Murakami:Jun2004}. We apply our theory to 4- and 6-band
Luttinger (inversion-symmetric with $\boldsymbol{\hat{P}}=0$) Hamiltonians
with anisotropic Luttinger parameters relevant to III-V semiconductor
compounds. The spherical model Hamiltonian in the presence of splitting
due to interactions with polarized Mn moments can be written as follows
within the mean-field description \citep{Abolfath:Jan2001}: \begin{equation}
\hat{H}_{0}=\dfrac{\hbar^{2}}{2m_e}\left[\left(\gamma_{1}+\dfrac{5}{2}\gamma_{2}\right)k^{2}-2\gamma_{2}(\mathbf{k}\cdot\hat{\mathbf{j}})^{2}\right]-\Omega\mathbf{m}\cdot\hat{\mathbf{s}}\;,\label{FourBand}\end{equation}
 where $\hat{\mathbf{j}}$ is the angular momentum operator for $J=3/2$,
$\hat{\mathbf{s}}$ is the spin operator which has to be projected
onto the $J=3/2$ total angular momentum subspace ($\hat{\mathbf{s}}=\hat{\mathbf{j}}/3$)
within the 4-band model, $\gamma_{1}$ and $\gamma_{2}$ are Luttinger
parameters defining the light- and heavy-hole bands with the effective
masses $m_{\rm lh/hh}=m_e/(\gamma_{1}\pm2\gamma_{2})$, in terms of the free-electron mass $m_e$, $\mathbf{m}$ is
the magnetization polarization direction and $\Omega$ is the mean
field proportional to the average of local moments. For fully-polarized
Mn spins, $\mathbf{m}$ is uniform and $\Omega=N_{\rm Mn}SJ_{\rm pd}$,
where $N_{\rm Mn}$ is the density of Mn ions with spin $S=5/2$,
and $J_{\rm pd}=50~{\rm meV}\,{\rm nm}^{3}$ is the strength of the exchange
coupling between the local moments and the valence-band electrons
\citep{Ohno:Oct1999}. The corresponding 6-band Hamiltonian
can be found in \citep{Abolfath:Jan2001}. As the vertex corrections
vanish, all terms involving $\mathbf{\hat{P}}$ in Eqs.~(\ref{Side-jump}) and (\ref{Skew scattering}) vanish, leading, up to linear
order in $\Omega$, to the following analytical result for Hamiltonian (\ref{FourBand}):\begin{equation}
\sigma^{\rm sj}_{yx}=\frac{\sigma_{0}}{10}\dfrac{5p(1-\sqrt{p})+3(1-p^{5/2})}{(1-p)(1+p-\sqrt{p})}\;,\label{Analytics}\end{equation}
 where $\sigma_{0}=(\Omega e^{2}/3\pi^{2}\hbar^{2})\sqrt{m_{\rm hh}/2\hbar\omega_{F}}$ and $p=m_{\rm lh}/m_{\rm hh}$. SJC is in the range from $0.3\sigma_{0}$ to $\sigma_{0}$ increasing as $p\rightarrow1$.
\begin{figure}
\includegraphics[clip,width=1\columnwidth]{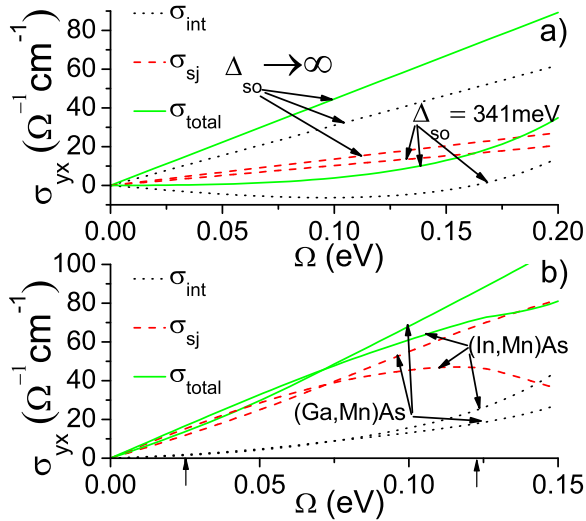} 

\caption{SJC and IC to the AHE as a function of the mean-field; (a) $\Delta_{\rm so}\rightarrow\infty$ corresponds to 4-band model
and $\Delta_{\rm so}=341~{\rm meV}$ corresponds to GaAs host, the hole
density is $0.35~{\rm nm}^{-3}$, (b) the plots correspond to the In/Ga,As hosts. The hole densities are $0.1~{\rm nm}^{-3}$ and $0.35~{\rm nm}^{-3}$ for the former/latter. By arrows, we mark the saturation mean-fields $\Omega=25/122~{\rm meV}$ for the In/Ga,As hosts \citep{Ohno:Oct1999,Jungwirth:may2002}.}

\label{fig2} 
\end{figure}

In Fig.~\ref{fig2}a, we present results of our calculations for
the spherical 4- and 6-band Hamiltonians. The parameters are chosen
to match GaAs effective masses $m_{\rm hh}=m_e/2$, $p=0.16$ and the
SO gap $\Delta_{\rm so}=341~{\rm meV}$. The SJC does not
change much as we vary $\Delta_{\rm so}$. The SJC can become dominant
for the smaller SO gaps since the IC sharply diminishes eventually
changing sign.

To have a more accurate description of the valence bands in III-V
semiconductor compounds, one has to introduce the third phenomenological
Luttinger parameter $\gamma_{3}$. This leads to band warping which
has strong effect on the IC \citep{Jungwirth:may2002}. Our calculations
show that the SJC is accelerated by the presence of band warping.
In Fig.~\ref{fig2}b, we present results of our calculations for
(In/Ga,Mn)As for which $\Delta_{\rm so}=430/341~{\rm meV}$, and $(\gamma_{1},\gamma_{2},\gamma_{3})=(19.67,8.37,9.29)/(6.85,2.1,2.9)$
\citep{Abolfath:Jan2001}. In both cases, the AHE is dominated by
the SJC. We use densities $N_{\rm Mn}=0.23/1.1~{\rm nm}^{-3}$
for the In/Ga,As host leading to saturation values of the effective
field $\Omega=25/122~{\rm meV}$ \citep{Ohno:Oct1999,Jungwirth:may2002}.
Taking the hole density as in Ref. \citep{Jungwirth:may2002} ($0.1/0.35~{\rm nm}^{-3}$
for In/Ga,As host), we arrive at the AHE $\sigma_{yx}=16/85~\Omega^{-1}\,{\rm cm}^{-1}$
for (In/Ga,Mn)As. Our results for the IC agree with the previous calculations
\citep{Jungwirth:may2002} while the total AHE overestimates the experimental
values \citep{Ohno:Oct1999,Edmonds:May2003} which is expected as
the experiments are only on the border of the metallic regime.

\textit{Summary.}---We formulated a theory of the AHE for metallic
noninteracting multi-band systems with the final result for the IC
and SJC being expressed through the material's electronic band structure. Our derivation relies on the minimal coupling with the electromagnetic field in Hamiltonian (\ref{Hamiltonian}), which is justified when a sufficient number of bands is considered. (E.g., the side-jump scattering in conduction bands due to spin-orbit coupling associated with impurities \citep{Nozieres:1973} can be described within our approach by resorting to the 8-band Kane model.) In contrast to the theory of the intrinsic AHE, the electron (hole) motion in a particular band cannot be defined solely in terms of the Bloch wave functions of the same band in the presence of disorder-induced band mixing. The SJC does not depend on the disorder strength but will generally change with the type of disorder (e.g., short-range impurities vs. phonon scattering).
The associated scattering regime crossovers can be accompanied by a sign change of the AHE as the IC and SJC can be of opposite sign. The AHE sign change has been observed in Fe and (Ga,Mn)As \citep{Dheer:Apr1967}. Ac measurements, furthermore, can quench
the SJC in clean samples at intermediate frequencies $\tau^{-1}<\omega<\Delta$.
We demonstrated our theory on electronic band structures of the 2D
Rashba and 3D Luttinger Hamiltonians. Within our simple model,
the AHE in the metallic (In/Ga,Mn)As magnetic semiconductors at low temperatures is dominated
by the SJC. The proposed theory can be further used in \textit{ab
initio} calculations of the AHE in wide range of available metallic
materials.

This work was supported in part by the DARPA, Alfred P. Sloan Foundation, NSF under Grant Nos. DMR-0840965 (YT) and DMR-0547875 (JS), and the Research Corporation for the Advancement of Science (JS).

\bibliographystyle{apsrev}

\end{document}